\documentclass{ws-ijmpa}

\usepackage[super,compress]{cite}
\usepackage{graphicx}
\usepackage{amsmath,amsfonts,amssymb,dsfont,bm,bbm,dcolumn,float,ifthen,mathrsfs,wasysym}
\usepackage{pstricks,feyn,verbatim,slashed,xcolor,multirow,array,rotating}
\usepackage{url,hyperref,fancybox,epstopdf}

\begin{document}

\markboth{Ligong Bian, Ning Chen, Yun Jiang}{Searches for Higgs Pairs in The CP-violating Two-Higgs-Doublet Model}

%
\catchline{}{}{}{}{}
%


\newcommand{\keV}   {~\mathrm{keV}}
\newcommand{\GeV}      {~\mathrm{GeV}}
\newcommand{\TeV}      {~\mathrm{TeV}}
\newcommand{\MeV}      {~\mathrm{MeV}}

\newcommand{\pb}      {~\mathrm{pb}}
\newcommand{\fb}      {~\mathrm{fb}}
\newcommand{\ab}      {~\mathrm{ab}}

\newcommand{\Tr}   {~\mathrm{Tr}}

\newcommand{\ba}{\begin{array}}
\newcommand{\ea}{\end{array}}
\newcommand{\beqn}{\begin{eqnarray}}
\newcommand{\eeqn}{\end{eqnarray}}
\newcommand{\beqs}{\begin{subequations}}
\newcommand{\eeqs}{\end{subequations}}
\newcommand{\be}{\begin{equation}}
\newcommand{\ee}{\end{equation}}
\newcommand{\non}{\nonumber \\}

\def\gU{\rm U}
\def\gSU{\rm SU}
\def\gO{\rm O}
\def\gSO{\rm SO}
\def\gSp{\rm Sp}
\def\gUSp{\rm USp}
\def\gE{\rm E}
\def\gF{\rm F}
\def\gG{\rm G}

\def\mA{\mathcal{A}}
\def\mB{\mathcal{B}}
\def\mC{\mathcal{C}}
\def\mD{\mathcal{D}}
\def\mE{\mathcal{E}}
\def\mF{\mathcal{F}}
\def\mG{\mathcal{G}}
\def\mH{\mathcal{H}}
\def\mI{\mathcal{I}}
\def\mJ{\mathcal{J}}
\def\mK{\mathcal{K}}
\def\mL{\mathcal{L}}
\def\mM{\mathcal{M}}
\def\mN{\mathcal{N}}
\def\mO{\mathcal{O}}
\def\mP{\mathcal{P}}
\def\mQ{\mathcal{Q}}
\def\mR{\mathcal{R}}
\def\mS{\mathcal{S}}
\def\mT{\mathcal{T}}
\def\mU{\mathcal{U}}
\def\mV{\mathcal{V}}
\def\mW{\mathcal{W}}
\def\mX{\mathcal{X}}
\def\mY{\mathcal{Y}}
\def\mZ{\mathcal{Z}}

\def\hf{\frac{1}{2}}

\newcommand{\vecMET}{\vec E\hspace{-0.08in}\slash_T}
\newcommand{\MET}{E\hspace{-0.08in}\slash_T}

\title{Higgs Pair Productions in the CP-violating Two-Higgs-Doublet Model}

\author{Ligong Bian}
\address{Department of Physics,\\
Chongqing University, Chongqing 401331, China\\
lgbycl@cqu.edu.cn}
\author{Ning Chen}
\address{Department of Physics, University of Science and Technology Beijing, \\
Beijing 100083, China\\
ustc0204.chenning@gmail.com}
\author{Yun Jiang}
\address{NBIA and Discovery Center, Niels Bohr Institute, University of Copenhagen, \\
Blegdamsvej 17, DK-2100, Copenhagen, Denmark\\
yunjiang@nbi.ku.dk}


\maketitle


\begin{abstract}

The SM-like Higgs pair productions are discussed in the framework of the general CP-violating two-Higgs-doublet model, where we find the CP-violating mixing angles can be related to the Higgs self couplings.
Therefore, the future experimental searches for Higgs boson pairs can be constrained by the improved precision of the electric dipole moment measurements. 
Based on a series constraints of the SM-like Higgs boson signal fits, the perturbative unitarity and stability bounds to the Higgs potential, and the most recent LHC searches for the heavy Higgs bosons, we suggest a set of benchmark models for the future high-energy collider searches for the Higgs pair productions.
The $e^+ e^-$ colliders operating at $\sqrt{s}= (500\,\GeV\,, 1\,\TeV)$ are capable of measuring the Higgs cubic self couplings of the benchmark models directly.
We also estimate the cross sections of the resonance contributions to the Higgs pair productions for the benchmark models at the future LHC and SppC/Fcc-hh runs.

\end{abstract}

\ccode{PACS numbers:12.60.Fr, 14.80.-j, 14.80.Ec}



\section{Introduction}
\label{section:introduction}

After the discovery of the $125\,\GeV$ Higgs boson at the LHC~\cite{Aad:2012tfa, Chatrchyan:2012xdj}, the most important process to unveil the underlying EWSB mechanism is through the direct measurements of the Higgs self couplings.
This can be done through the Higgs pair productions at both high-energy $e^+ e^-$ and $pp$ colliders.
The current LHC searches for the Higgs pair productions focus on the leading production channel of gluon-gluon fusion (ggF). 
From the experimental side, it is well-known that several future high-energy collider programs, such as the International Linear Collider (ILC)~\cite{Baer:2013cma} in Japan, the Future eplus-eminus/hadron-hadron Cicular Collider (Fcc-ee/Fcc-hh)~\cite{Gomez-Ceballos:2013zzn} at CERN, and the Circular electron-positron Collider (CEPC)/ Super-$pp$-Collider(SppC)~\cite{CEPC-SPPC-pre} in China, have been proposed in recent years.

The CP-violation (CPV) 2HDM is likely to realize the EW baryogenesis~\cite{Bian:2014zka}, which is one of the most popular solutions to the baryon asymmetry in the Universe.
Wherein, the $125\,\GeV$ SM-like Higgs boson, often chosen to be $h_1$, is a CP mixture~\cite{Lavoura:1994fv, Barroso:2012wz, Brod:2013cka,Mao:2014oya,Mao:2016jor}.
Thus, the CPV couplings for the SM-like Higgs bosons are subject to the constraints from the searches for the electric dipole moments (EDMs).~\footnote{See, e.g., Refs.~\cite{Pospelov:2005pr,Engel:2013lsa} for recent reviews. }
Together with other existing constraints to the CPV 2HDM, 
one can find the constraints to the heavy Higgs boson mass ranges and the Higgs cubic self couplings.
Hence, the cross sections of the Higgs pair productions in the CPV 2HDM can be predicted at the future $e^+ e^-$ and $pp$ colliders.

Here, we study the Higgs pair productions in the framework of the CPV 2HDM, including the precise measurement of the SM-like Higgs cubic self couplings at the $e^+ e^-$ colliders, and the resonance contributions in the gluon-gluon fusion (ggF) production channel at the $pp$ colliders.
In Sec.~\ref{section:CPV2HDM}, we review the setup of the CPV 2HDM.
%
In Sec.~\ref{section:constraint}, we impose constraints to the CPV 2HDM-II parameter space.
The main results of the Higgs pair productions in the CPV 2HDM are presented in Sec.~\ref{section:collider}.
By combining the current constraints, 
a set of benchmark models are given.
We estimate the physical opportunities of the precise measurement of the SM-like Higgs cubic self coupling $\lambda_{111}$ at the future high-energy $e^+ e^-$ colliders, with focus on the $e^+ e^- \to hhZ$ process at the $\sqrt{s}=500\,\GeV$ run.
The heavy resonance contributions to the Higgs pair productions can become dominant at the $pp$ colliders.
The conclusions and discussions are given in Sec.~\ref{section:conclusion}.


\section{The CPV 2HDM}
\label{section:CPV2HDM}

\subsection{The CPV 2HDM potential}

A constraint between mixing angles and mass eigenvalues of neutral Higgs bosons are given as follows~\cite{Khater:2003wq}
\beqn\label{eq:mass_constraint}
&&( M_1^2 - M_2^2 s_{\alpha_c}^2  - M_3^2 c_{\alpha_c}^2 ) s_{\alpha_b} (1 + t_\alpha)= (M_2^2 - M_3^2 ) (t_\alpha t_\beta - 1) s_{\alpha_c} c_{\alpha_c}\,,
\eeqn
with $M_1=125\,\GeV$ assumed.
The parameter inputs are simplified by requiring all heavy Higgs boson masses are degenerate, i.e., $M_{2}=M_{3}=M_\pm \equiv M$. 
This was usually taken to relax the constraints from the electroweak precision measurements. 
The constraint of Eq.~\eqref{eq:mass_constraint} among the mixing angles becomes
$\alpha_b=0 \, ~\textrm{or}~ t_\alpha=-1\,.$
Below, we will always take $\alpha=-\pi/4$.~\footnote{
The study of the phenomenology with the CPV mixings of $|\alpha_b|\ll |\alpha_c|$ is carried out in a separate work~\cite{Bian:2016zba}.
}
Without loss of generality, we always take $\alpha_c=0$ for simplicity.
Thus, the set of input parameters can be summarized as follows
\beqn\label{eq:inputs}
&&M_1=125\,\GeV\,, \qquad M_2=M_3=M_\pm = M\,, \qquad m_{\rm soft} \non
&&\alpha=-\frac{\pi}{4}\,,\qquad  t_\beta \,, \qquad  \alpha_b\,,\qquad \alpha_c=0 \,.
\eeqn
%


\section{The Constraints in The CPV 2HDM}
\label{section:constraint}

The ACME experiment~\cite{Baron:2013eja},which searches for an energy shift of ThO molecules due to an external electric field, set stringent experimental bound to the eEDM as $|d_e/e| < 8.7\times 10^{-29}\, {\rm cm} $. 
In the CPV 2HDM, the EDM $d_e$ are contributed by the two-loop Barr-Zee type $h_i \gamma\gamma$($h_i Z \gamma$) diagrams~\cite{Barr:1990vd}, and the $H^\pm W^\mp \gamma$ diagrams. 
The combined $125\,\GeV$ Higgs boson signal constraints and the eEDM constraints for the CPV 2HDM-II
allow region of the CPV mixing angle up to $|\alpha_b |\lesssim 0.1$, while the $1\,\sigma$ allowed range of $t_\beta$ is basically around $1.0$~\cite{Bian:2016awe}. 
In order to highlight the CPV effects in the Higgs self couplings in the following discussions, we will focus on the CPV 2HDM-II with the fixed inputs of $\alpha=-\pi/4$ and $t_\beta=1.0$.

To have a self-consistent description of the 2HDM potential, two other theoretical constraints should be taken into account, namely, the perturbative unitarity and the stability~\cite{Arhrib:2000is, Kanemura:2015ska}.
We shall also take into account the constraints from the $7\oplus 8\,\TeV$ LHC searches for the heavy Higgs bosons in the 2HDM spectrum. 
Combining with the unitarity and stability constraints, 
 we consider two scenarios of benchmark models for the $|\alpha_b|=0.1$ and $|\alpha_b|=0.05$ cases~\cite{Bian:2016awe}.

\section{Higgs Pair Productions at The Colliders}
\label{section:collider}

In this section, we study the SM-like Higgs pair productions in the framework of the CPV 2HDM. 

\subsection{The precise measurement of $\lambda_{111}$ at the future $e^+ e^-$ colliders}

\begin{figure}
\centering
\includegraphics[width=5cm,height=4cm]{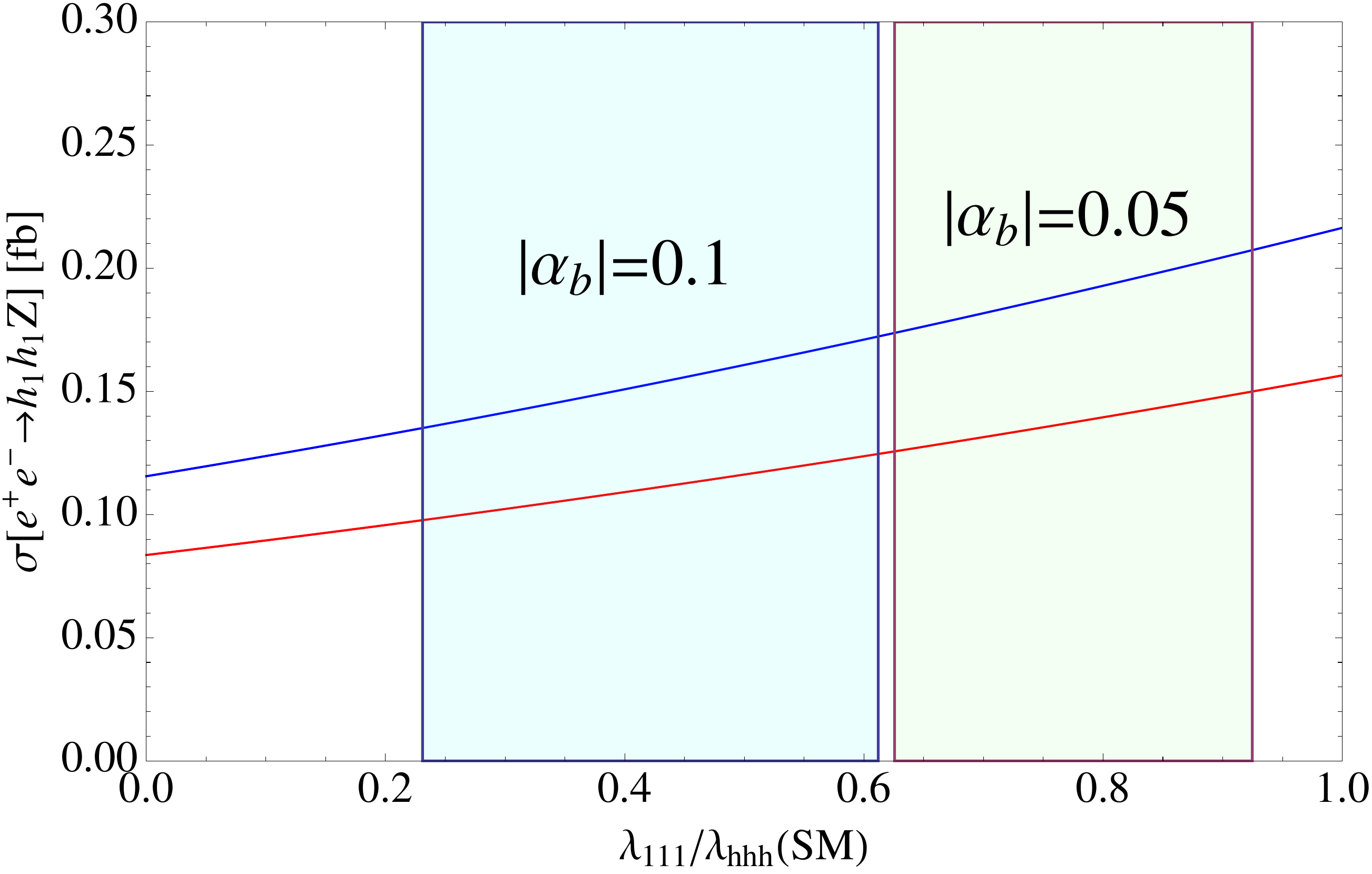}
\includegraphics[width=5cm,height=4cm]{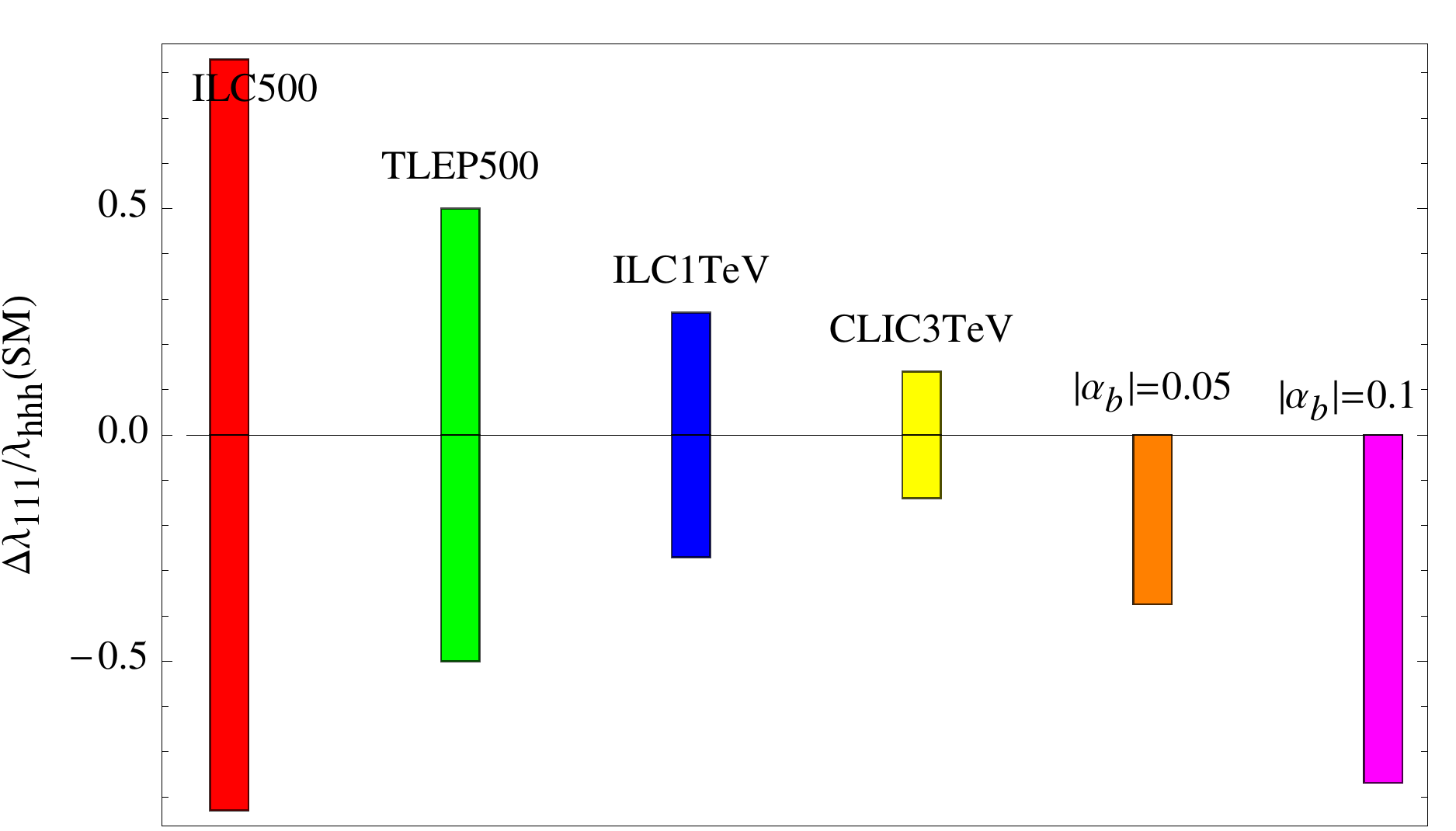}
\caption{\label{fig:eehhZ} 
Left: the cross sections of $\sigma[e^+ e^- \to h_1 h_1 Z]$ at the TLEP (red) and ILC (blue) $500\,\GeV$ versus the different Higgs cubic self couplings.
Right: the expected accuracies on the Higgs cubic self couplings at the future $e^+ e^-$ colliders, and the $\Delta \lambda_{111}/\lambda_{hhh}^{\rm SM}$ for the benchmark models of $|\alpha_b|=0.1$ and $|\alpha_b|=0.05$.
}
\end{figure}

The direct measurements of the Higgs self couplings can be achieved via the $e^+ e^- \to h h Z$ process with the center-of-mass energy of $\sqrt{s} = 500\,\GeV$~\cite{Baer:2013cma,Gomez-Ceballos:2013zzn,Tian:2010np}.
%
%
The ratio of the total cross section of $\sigma[e^+ e^- \to h h Z]$ to its SM counterpart can be parametrized as follows
\beqn
\frac{\sigma[e^+ e^- \to h_1 h_1 Z]}{~~~\sigma[e^+ e^- \to h h Z]_{\rm SM}} &=& 0.097\,\xi_{111}^2 + 0.369\,\xi_{111} + 0.534\,,
\eeqn
at the TLEP and ILC $500\,\GeV$ runs, with $\xi_{111} \equiv \lambda_{111}/\lambda_{hhh}^{\rm SM}$.
The total cross sections at the TLEP and ILC $500\,\GeV$ runs versus the ratios of different Higgs cubic self couplings $\lambda_{111}/\lambda_{hhh}^{\rm SM}$ are displayed on the left panel of Fig.~\ref{fig:eehhZ}.
On the right panel of Fig.~\ref{fig:eehhZ}, we display the expected accuracies on the Higgs cubic self couplings for ILC500 (with $\int \mL dt=0.5\,\ab^{-1}$), TLEP500 (with $\int \mL dt=1\,\ab^{-1}$), ILC $1\,\TeV$ (with $\int \mL dt=1\,\ab^{-1}$), and CLIC $3\,\TeV$ (with $\int \mL dt=2\,\ab^{-1}$).
For the $|\alpha_b|=0.1$ case, the largest deviations of $\lambda_{111}$ can be probed with the accuracies reached by the TLEP $500\,\GeV$; while for the smaller CPV mixing angle of $|\alpha_b|=0.05$ case, the largest deviations of $\lambda_{111}$ can be probed by the ILC $1\,\TeV$.

\subsection{The $p p\to h_1 h_1$ in the CPV 2HDM}

Now we present the results of the Higgs pair productions in the CPV 2HDM based on all previous constraints.
The cross sections are obtianed by using the FeynRules~\cite{Christensen:2008py} for model implementation and Madgraph 5~\cite{Alwall:2014hca}.
From the previous estimation of the Higgs cubic self couplings for the $M_2=M_3=600\,\GeV$ case, we may either have the large resonance contributions or go to the regions with the vanishing resonance contributions of $(\lambda_{111}\,, \lambda_{113})\to (\lambda_{hhh}^{\rm SM}\,, 0)$.
%
%

\begin{figure}
\centering
\includegraphics[width=5cm,height=4cm]{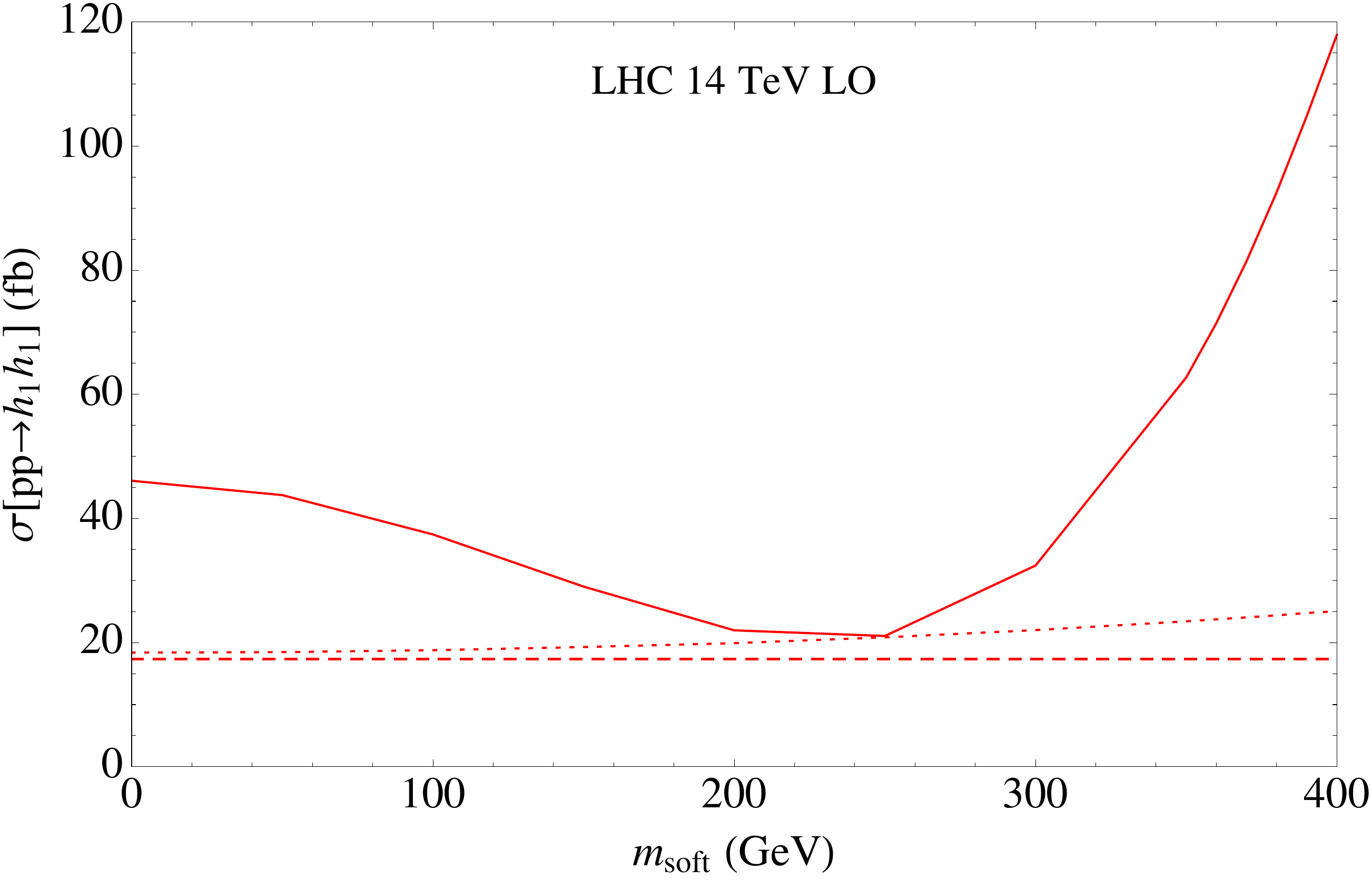}
\includegraphics[width=5cm,height=4cm]{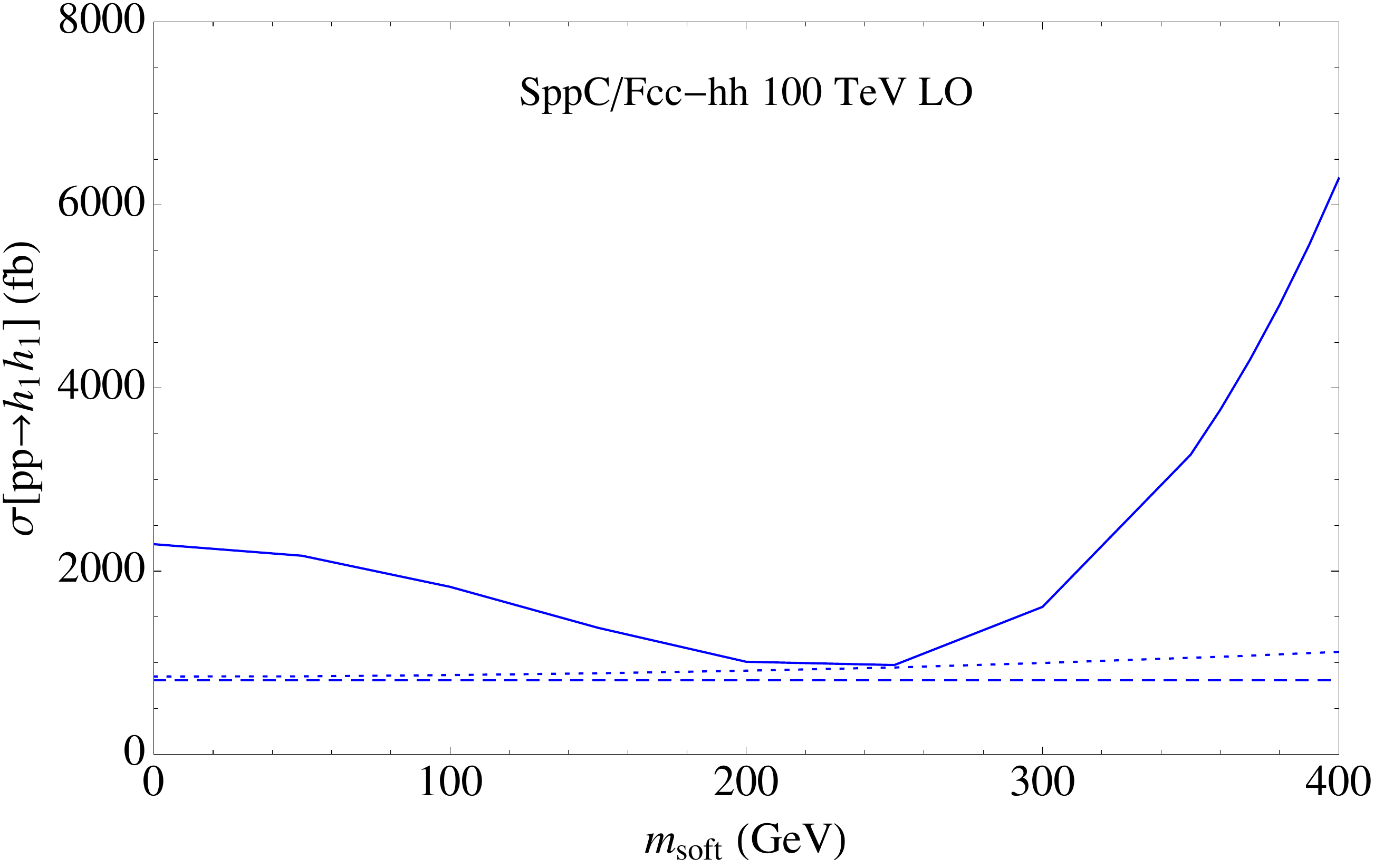}
\caption{\label{fig:sigmas} 
The cross sections of $\sigma[pp\to h_1 h_1]$ at the LHC $14\,\TeV$ (left) and SppC $100\,\TeV$ (right) versus the varying $m_{\rm soft}$ for the $M_2=M_3=600\,\GeV$ case in the CPV 2HDM-II, with fixed inputs of $|\alpha_b|=0.1$. 
}
\end{figure}

In Fig.~\ref{fig:sigmas}, we display the LO cross sections of $\sigma[pp\to h_1 h_1]$ at the LHC and the SppC/Fcc-hh for the $M_2=M_3=600\,\GeV$ case.
The solid curves represent the total cross sections.
We also show the hypothetical cross sections by dotted curves, where we turn off the Higgs cubic self coupling of $\lambda_{113}$ and modify $\lambda_{111}$.
Thus, it is evident that the total cross sections approach to the SM-like Higgs pair productions with the modified cubic self couplings.
On the other hand, the LO cross sections at the LHC (SppC) can be as large as $\sim\mO(100)\,\fb$ ($\sim \mO(6)\,\pb$) when the soft mass approaches to the stability boundary for this case.

\section{Conclusion and Discussion}
\label{section:conclusion}

In this work, we study the Higgs pair productions in the framework of the CPV 2HDM-II by imposing theoretical and experimental constraints.
The Higgs cubic self couplings play the most crucial role for the Higgs pair production.
For our case, two relevant cubic self couplings are $\lambda_{111}$ and $\lambda_{113}$, which are controlled by the soft mass term $m_{\rm soft}$ and the CPV mixing angle of $\alpha_b$.
The precise measurement of the SM-like Higgs cubic coupling of $\lambda_{111}$ can be achieved via the $e^+ e^- \to h_1 h_1 Z$ process at the future high-energy $e^+ e^-$ colliders.
The benchmark models in our discussions typically predict totally cross sections of $\sigma[e^+ e^- \to h_1 h_1 Z]$ smaller than the SM predictions.
The largest deviations of the SM-like Higgs cubic couplings $\lambda_{111}$ are likely to be probed at the future TLEP $500\,\GeV$ and ILC $1\,\TeV$ runs.
At the future high-energy $pp$ collider runs, the Higgs pair productions are very likely to be controlled by the heavy resonance contributions. 
In the allowed mass range of the heavy Higgs bosons, we find the total production cross sections to be $\sigma[pp\to h_1 h_1]\sim \mO(10)- \mO(100)\,\fb$ at the LHC 14 TeV runs.
They can be as large as $\sim \mO(10^3)\,\fb$ at the future SppC 100 TeV runs.
The discovery of these channels will manifest the structure of the Higgs sector.
Therefore, it will be very helpful to further study the higher-order QCD corrections as well as the collider search capabilities for such heavy resonance contributions to the Higgs pairs.


\section*{Acknowledgments}

The work of LGB and NC are supported by the National Science Foundation of China under grant No. 11605016 and No. 11575176. 
Y.J. acknowledges financial support by the the HKIAS where this work was presented during the IAS Program on High Energy Physics Conference.
We would like to thank the conference organizers for the invitation, and the members of the Jockey Club Institute for Advanced Study for their hospitality.


\end{document}